\documentclass[aps,pre,twocolumn,superscriptaddress,showpacs]{revtex4}
\usepackage{graphicx}
\usepackage{amsmath}

\begin{document}


\title{
Origin of rebounds with a restitution coefficient larger than unity in 
nanocluster collisions
}


\author{Hiroto Kuninaka}
\email[E-mail: ]{kuninaka@edu.mie-u.ac.jp}
\affiliation{Faculty of Education, Mie University, 
Tsu city, Mie, Japan, 514-8507}
\author{Hisao Hayakawa}
\affiliation{Yukawa Institute for Theoretical Physics, 
Kyoto University, Sakyo-ku, Kyoto, Japan, 606-8502
 }

\date{\today}

\begin{abstract}
We numerically investigate the mechanism of super rebounds 
for head-on collisions between nanoclusters  
in which the restitution coefficient is larger than unity. 
It is confirmed that the temperature and the entropy of the nanocluters 
decrease after the super rebounds 
by our molecular dynamics simulations. It is also found
that the initial metastable structure plays a key role for the
emergence of the super rebounds.
\end{abstract}


\maketitle
\section{Introduction}
Nanoclusters are technologically important for the construction of nanodevices. 
Because the size of nanoclusters is mesoscopic, 
thermodynamic properties of such materials are still not 
well understood\cite{hill}, 
though the methods to make nanoclusters such as adiabatic expansion 
through a nozzle and a laser ablation technique  are well established.\cite{baletto}

Dynamics of nanoclusters are extensively investigated 
from both scientific and technological interest. 
There are many numerical studies on cluster-cluster and cluster-surface 
collisions based on the molecular dynamics simulation.
\cite{awasthi2010,awasthi2010-2,han,kuninaka_hayakawa_PRE2009,
saitoh_prog,suri,antony,awasthi,kalweit06,kalweit,tomsic,yamaguchi,
harbich,valentini,knopse,ming}  
We observe variety of rebound processes for such systems 
caused by the competition between  
the attractive interaction and the repulsive interaction 
of two colliding bodies. 
Binary collisions of identical clusters cause 
coalescence, scattering, and fragmentation 
depending on the cluster size and the impact energy.
\cite{kalweit06,kalweit} 
On the other hand, cluster-surface collisions induce 
soft landing, embedding, and fragmentation.\cite{saitoh_prog, harbich, valentini}
The attractive interaction plays crucially important roles 
in collisions of nanoclusters, so that the modeling of the cohesive collisions 
in various scale are actively discussed in these days.\cite{brilliantov07,kim,kosinki} .

However, the attractive interaction can be reduced by some 
combinations of the two interacting objects and the relative 
configuration of colliding molecules.\cite{sakiyama} 
Awasthi {\it et al.} introduced a modified Lennard-Jones model 
to simulate the rebound process of a $\rm{Bi}$ cluster onto a 
$\rm{SiO_{2}}$ surface\cite{awasthi}, in which  
they introduced a cohesive parameter to reduce 
the attractive interaction between different atoms on the surface. 
There exists a corresponding experiment on the 
rebound process of Bi nanoclusters on a Si surface 
with the aid of a V-shaped template etched on 
a silicon wafer.\cite{awasthi2010} 
This suggests that the modified Lennard-Jones clusters can approximately 
describe collisions of real clusters. 
Similarly, recent papers have reported 
that surface-passivated Si nanoclusters exhibit elastic rebounds on 
Si surface due to the reduction of the attractive interaction 
between the surfaces.\cite{suri,hawa,saitoh} 
In particular, Saitoh {\it et al.} confirmed that 
the behavior in collisions of modified Lennard-Jones 
clusters is similar to that of H-passivated Si clusters from their simulation.\cite{saitoh} 
These results also support that the modified Lennard-Jones model can be regarded 
as a simplified model of nanoclusters.

In general, thermal fluctuation also plays an important role 
for small systems such as nanoclusters. 
Indeed, the present authors carried out 
the molecular dynamics simulation of colliding thermally activated 
modified Lennard-Jones clusters to investigate the impact phenomena 
and found the existence of the ``super-rebound" in which 
the restitution coefficient is larger than unity.\cite{kuninaka_hayakawa_PRE2009} 
Recently, it has been reported that such a rebound can be 
observed in a molecular dynamics simulation 
of a collision of Cu nanoparticles on a rigid wall.\cite{han} 
In addition, another research group has reported that large recovery strain 
after loading (called super-elasticity) can be found in a small system.\cite{Kenichi_Saitoh} 
Although there is a possibility that such an anomalous mechanical property of 
small systems is also concerned with the emergence of  ``super rebounds", 
the underlying mechanism for the ``super rebounds" is still unclear. 

In macroscopic systems, 
the restitution coefficient larger than unity can be observed in 
oblique collisions between a hard sphere and a soft elastic plate
\cite{louge,kuninaka_hayakawa_PRL2004}, 
which differs from the super rebound. Indeed, the restitution
coefficient can easily become larger than unity if the rebound direction
is changed in oblique collisions. As another example, 
a recent study reported that the large-scale magnetized plasmoids can 
gain the increased kinetic energy 
after collision in the heliosphere.\cite{shen} 
On the contrary, in microscopic systems such as nanoclusters, 
the true ``super rebounds" can be observed in normal collisions, 
which implies that a part of thermal energy is converted 
to macroscopic degrees of freedom. 
This may imply a possibility to make a nanoscale object which extracts work 
from its internal energy.  
Although the averaged behavior of the restitution coefficient 
against the relative impact speed can be approximately understood  
by the macroscopic theory of cohesive collisions,\cite{brilliantov07,kuninaka_hayakawa_PRE2009} 
the mechanism of curious energy transfer against intuition of the thermodynamics 
is important  to both fundamental physics and energy technology. 

The aim of this paper is to study the mechanism of the 
super rebounds from the viewpoint of the energy transfer between 
the microscopic and the macroscopic degrees of freedom. 
For this purpose, we investigate the characteristics of  thermodynamic 
functions such as temperature and entropy in the super rebounds 
based on the molecular dynamics simulation. 
In addition, we also investigate the change in structures of 
the colliding nanoclusters characterizing super rebounds 
by the introduction of Steinhart's order parameter\cite{steinhardt} and some 
related geometrical order parameters.

The organization of this paper is as follows. In the next section, 
we introduce our nanocluster model. In Sec. III we show the results of 
our molecular dynamics simulations. 
In Sec. IV we discuss the reason why super rebounds can be 
observed in our simulation. Section V is devoted to the summary of this work. 
In Appendices A and B, we explain the method of calculating entropy 
in our simulation and how to calculate Steinhardt's order parameters, 
respectively.

\section{Model}
\begin{figure}[hbtp]
\begin{center}
\includegraphics[width=.3\textwidth]{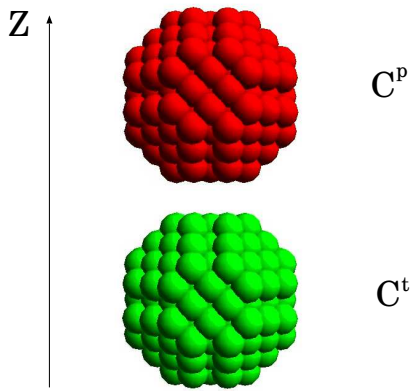}
\end{center}
\caption{
(Color online) Snapshot of our model after equilibration to 
$T=0.02\epsilon$. 
Each of them contains 236 ``atoms'' which are bound together 
by the modified Lennard-Jones potential.
}
\label{nc001}
\end{figure}

Let us introduce our model. Our model consists of two identical 
nanoclusters, each of which contains $236$ ``atoms'' (Fig. \ref{nc001}). 
The clusters have facets due to the small number of ``atoms''. 
To construct one cluster, we first make the face-centered cubic crystal 
of $9 \times 9 \times 9$ layers of atoms.
Next, we cut out a spherical cluster from the cube. 
In Fig.\ref{nc001}, we call the upper projectile and 
the lower target clusters C$^{p}$ and C$^{t}$, respectively. 
All the atoms in one cluster are bound together 
by the modified Lennard-Jones potential 
\begin{equation}\label{LJ}
U(r_{ij})=4\epsilon\left\{
\left(\frac{\sigma}{r_{ij}}\right)^{12}-
c \left(\frac{\sigma}{r_{ij}}\right)^{6}
\right\}, 
\end{equation}
where $r_{ij}$ is the distance between two arbitrary atoms 
$i$ and $j$ in this system, $\sigma$ and $\epsilon$ are the core diameter 
and the scale of energy, respectively. 
Here, $c$ is the cohesive parameter which changes 
the magnitude of cohesion between atoms. In our simulation,  
we adopt $c=0.2$ between the atom $i$ on the lower surface of C$^{p}$ 
and the atom $j$ on the upper surface of C$^{t}$ while $c=1$ between 
all the atoms in each cluster. 
Here the definition of the surface and the bulk atoms is as follows. 
Each cluster consists of $9$ layers of atoms. 
We define the atoms outside the concentric inscribed sphere 
as the surface atoms while the other atoms 
as the bulk atoms. 
We assume that the surface atoms are different from the bulk atoms, 
which may cause the weak cohesive interaction 
between the clusters.\cite{awasthi,sakiyama} 

Let us introduce the volume fraction $\varphi$ which is defined by
\begin{equation}
\varphi \equiv \frac{1}{6}\pi \rho \sigma^{3}, 
\end{equation}
where $\rho$  is the number density. 
\begin{figure}[hbtp]
\begin{center}
\includegraphics[width=.4\textwidth]{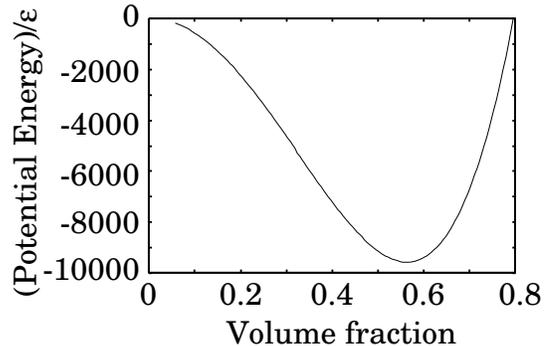}
\end{center}
\caption{
Relationship between volume fraction of atoms 
and potential energy of one cluster.
}
\label{pot}
\end{figure}
Figure \ref{pot} is the relationship between $\varphi$ and 
the initial potential energy of one cluster. To construct a cluster, 
we adopt the value for $\varphi$ as $\varphi=0.56$ in which the potential energy is minimum, 
while we adopted $\varphi=0.21$ in our previous work.
\cite{kuninaka_hayakawa_PRE2009}

The procedure of our simulation is as follows. 
As the initial condition of the simulation, 
the centers of mass of C$^{p}$ and C$^{t}$ are placed 
along the $z-$axis with the separation $4R$ between the centers of mass  
of C$^{p}$ and C$^{t}$, 
where $R$ is the radius of 
the cluster (Fig. \ref{nc001}). The two clusters are placed in mirror symmetric positions  
with respect to $z=0$ so that a facet of one cluster is placed face-to-face with 
that of another cluster. We have checked that the relative rotational orientation 
of the clusters around $z$-axis little affect the relationship 
between the restitution coefficient and the colliding speed in our previous work. 
\cite{kuninaka_hayakawa_PRE2009}
The initial velocities of the atoms in both C$^{p}$ and C$^{t}$ 
satisfies the Maxwell-Boltzmann distribution with the initial temperature $T$. 
The sample average is taken over different sets of initial velocities.

We numerically solve 
the equation of motion of each atom $i$ described by
\begin{eqnarray}\label{eqm}
M \frac{d^{2} {\bf x}_{i}}{dt^{2}} = \sum_{j \ne i} {\bf F}_{ij} 
+ \sum_{k} {\bf F}_{ik}. 
\end{eqnarray}
where $M$ and ${\bf x}_{i}$ are the mass of an atom and the position of 
the $i$-th atom, respectively. ${\bf F}_{ij}$ is 
the modified Lennard-Jones force  which is calculated from Eq. (\ref{LJ}) as 
\begin{equation}
{\bf F}_{ij} = 
-\frac{\partial U(r_{ij})}{\partial {\bf r}_{ij}}.
\end{equation}
Numerical integration of the equation of motion for each atom 
is carried out by the second order symplectic integrator with 
the time step $dt=1.0 \times 10^{-2} \sigma/\sqrt{\epsilon/M}$. 
To reduce computational costs, we introduce the cut-off length $r_{c}$ 
of the Lennard-Jones interaction as $r_{c}=2.5 \sigma$. 
The rate of energy conservation, $|E(t)-E_{0}|/|E_{0}|$, 
is kept within $10^{-5}$, 
where $E_{0}$ is the initial energy of the system and $E(t)$ is the 
energy at time $t$.

We equilibrate the clusters by using the velocity scaling method 
\cite{woodcock,nose} as the thermostat 
in the initial $2000$ simulation steps. 
Here, we introduce the kinetic temperature, 
\begin{equation}\label{temp}
T = \frac{2}{3N} \sum_{i} \frac{1}{2} M ({\bf v}_{i}-{\bf v}_{c})^{2},
\end{equation}
where $N$ and ${\bf v}_{c}$ are the number of atom and the velocity of 
the center of mass of one cluster, respectively. 
It has been confirmed that the temperature of the system 
converges on a desired temperature during the equilibration.


After the equilibration we give the translational velocity to C$^{p}$ 
and C$^{t}$ to make them collide against each other without the thermostat. 
We give the translational speed with the acceleration 
$g=0.02 \epsilon/(\sigma M)$. 
The typical value of the relative impact speed in our simulation 
is $V = 0.1\sqrt{\epsilon/M}$, 
which is slightly less than the thermal velocity 
of the system defined by $V_{\rm{th}}=\sqrt{T/M}$ when $T=0.02\epsilon$. 


\section{Simulation Results}\label{modelA}
\subsection{Macroscopic Properties}



Here we show the macroscopic properties of the colliding nanoclusters 
in our simulation. 
To characterize the rebound behavior of macroscopic inelastic collisions, 
we calculate the restitution coefficient $e$ defined by 
\begin{equation}
e = \frac{V_{z}(\bar{t}^{*})}{V_{z}(0)},
\end{equation}
where $V_{z}(\bar{t}^{*})$ is the $z$ component of 
the relative translational speed of C$^{p}$ to C$^{t}$, 
and $\bar{t}^{*}$ is the scaled time of separation of the 
clusters, $\bar{t}^{*} \equiv t^{*}/(\sigma/\sqrt{\epsilon/M})$. 
We define the time of separation $t^{*}$ by the time when 
the relative rebound speed becomes stable after the collision.

\begin{figure}[hbtp]
\begin{center}
\includegraphics[width=.4\textwidth]{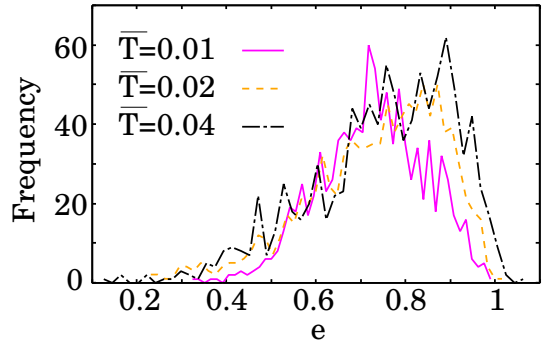}
\end{center}
\caption{
(Color online)
Histogram of $e$ for several temperatures of clusters with 
$V_{z}=0.2 \sqrt{\epsilon/M}$. 
}
\label{V0_2}
\end{figure}

Figure \ref{V0_2} shows the histograms of $e$ for initial temperatures  
$T=0.01\epsilon, 0.02\epsilon,$ and $0.04\epsilon$, respectively. 
Each histogram is constructed from 1000 samples with the initial speed of 
$V_{z} = 0.2 \sqrt{\epsilon/M}$. We find $12$ samples of the super rebound 
at $T=0.04\epsilon$ 
while all samples 
are the ordinary rebounds at $T=0.01\epsilon$. At $T=0.02\epsilon$ 
we found that only 1 sample is the super rebound. 
On the other hand, if we adopt $V_z=0.3\sqrt{\epsilon/M}$, most
of the rebounds are ordinary. We can find only two samples of super
rebounds even at $T=0.04\epsilon$.
Thus, the super rebounds can be observed only if the colliding speed is lower or 
almost equal to the thermal speed.

\subsection{Thermodynamic Properties}

\begin{figure}[hbtp]
\begin{center}
\includegraphics[width=.4\textwidth]{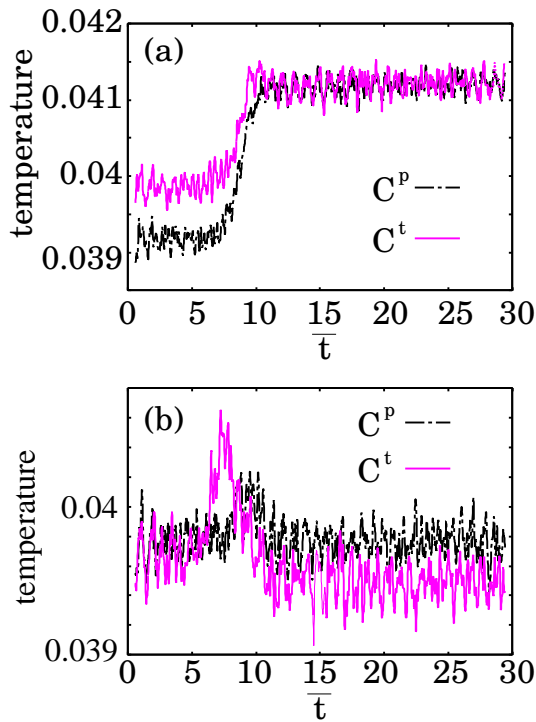}
\end{center}
\caption{
(Color online)
Time evolution of temperature for (a) an ordinary rebound 
and (b) a super rebound. 
}
\label{temp_modelB}
\end{figure}

Here we show the time evolution of kinetic temperature defined by Eq.(\ref{temp}). 
Figure \ref{temp_modelB} shows typical examples of the changes in temperature 
in an ordinary rebound and a super rebound. 
$\bar{t}$ is the scaled time, $\bar{t} \equiv t/(\sigma/\sqrt{\epsilon/M})$. 
A slight discrepancy in temperature between 
C$^{p}$ and C$^{t}$ is observed after the equilibration in Fig. \ref{temp_modelB} (a). 
The temperature increases after the collision around $\bar{t}=9$ 
for an ordinary rebound (see Fig. \ref{temp_modelB}(a)), 
where a part of the translational energy is converted to the internal energy. 
On the other hand, the temperature decreases in C$^{t}$ while the temperature 
of C$^{p}$ increases after the collision for a super rebound (Fig. \ref{temp_modelB}(b)). 


Next let us investigate the entropy change $\Delta S$ defined by Eq. (\ref{entropy}) 
in Appendix \ref{appA}. 
Figure \ref{ent_modelB}(a) plots the time
evolution of $\Delta S$ for C$^{p}$ and C$^{t}$ in a super rebound, which
shows that the entropy increases in C$^{t}$ while it decreases
in C$^{p}$ during collision for $\bar{t}<15.8$. On the other hand, 
Figure \ref{ent_modelB}(b)
shows that the discrepancy of entropy change remains
finite after the collision for $\bar{t}>15.8$. In addition, it is remarkable
that the entropy of C$^{t}$ decreases corresponding to the decrease of temperature.

So far, we have reported the results of our simulation on thermodynamic
quantities. What we confirm is that the behavior of such quantities 
in super rebounds is seemingly in contrast to what is expected 
from the second law of thermodynamics. 
However, we should note that the second law is only strictly valid 
after ensemble average is taken over. In this sense, our result does 
not violate the second law of thermodynamics.

\begin{figure}[t]
\begin{center}
\includegraphics[width=.38\textwidth]{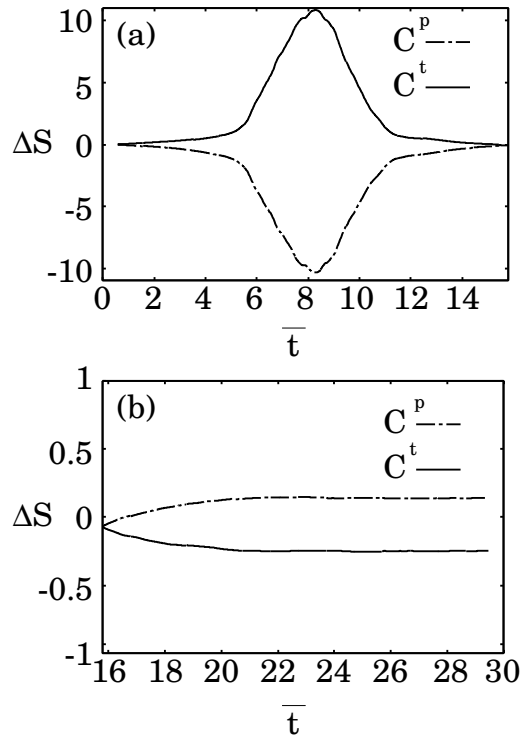}
\end{center}
\caption{
Time evolution of $\Delta S$ of a super rebound in (a) $\bar{t} \le 15.8$ 
and (b) $\bar{t} \ge 15.8$. 
Chain and solid lines show $\Delta S$ in C$^{p}$ and C$^{t}$, respectively. 
At $\bar{t}=30$, $\Delta S$ of C$^{t}$ becomes $-0.243$. }
\label{ent_modelB}
\end{figure}

\subsection{Structural Change during Collision}\label{structB}
Let us investigate the structural change of the clusters induced by collisions. 
To characterize the structural change, we first introduce 
a local bond order parameter known as time-averaged Steinhart's bond order 
parameter $Q_{l}$ 
(see Eq.(\ref{Qli}) for its explicit definition).
\cite{steinhardt,lechner,wolde} 
We note that 
Steinhardt's bond order parameters are used to characterize the structural 
change in nanoclusters associated with collisions\cite{jung} and 
melting\cite{suka,wang} as well as equilibrium structures of 
crystalline solids. 
According to the definition of the bond order parameter 
summarized in Appendix \ref{appB}, we calculate $Q_{4}(i)$ and $Q_{6}(i)$ 
of each atom $i$ before and after collisions, respectively.  
For the time average in calculation of $Q_{4}(i)$ and $Q_{6}(i)$, 
we use $t_{b}=5.5 \sigma/\sqrt{\epsilon/M}$ 
and $t_{a}=20\sigma/\sqrt{\epsilon/M}$ for before and after collisions, 
respectively, with the time interval 
$\tau_{\alpha}=1.5\sigma/\sqrt{\epsilon/M}$, 
where $t_a$ and $t_b$ respectively correspond to $t_0$ in Eq. (\ref{eqB3}). 

On $Q_{4}$-$Q_{6}$ plane, the highest peak can be found 
around $(Q_{4}, Q_{6})=(0.190, 0.574)$ which is characteristic for FCC crystal structure. 
We cannot find visible shift of the peak position on $Q_4$-$Q_6$ plane 
during the collision in both super and ordinary rebounds.  
However, we find that the number of ``atoms'' at the highest peak decreases after the collision 
for ordinary rebounds, which means that the number of ``atoms'' in FCC bond order 
decreases after the collision (see Table \ref{tabl_ord}). 
On the contrary, for super rebounds, the number of ``atoms'' at the highest peak 
increases after the collision, 
which means that the FCC bond order becomes intensive after the collision. 
Thus, the super rebounds are characterized by the increase of the 
number of ``atoms'' with FCC bond order during collisions.

 \begin{table}
 \caption{\label{tabl_ord}
Numbers of FCC ``atoms'' for ordinary rebound.
}
 \begin{center}
 \begin{tabular}{lccc}
 & numbers of FCC ``atoms'' & $Q_{4}$ & $Q_{6}$\\
\hline
before collision & $649$ & $0.19105$ & $0.56897$\\
after collision & $574$ & $0.1908$ & $0.5689$\\
 \end{tabular}
 \end{center}
 \end{table}

 \begin{table}
 \caption{\label{tabl_sup}
Numbers of FCC ``atoms'' for super rebound.
}
 \begin{center}
 \begin{tabular}{lccc}
 & numbers of FCC ``atoms'' & $Q_{4}$ & $Q_{6}$\\
\hline
before collision & $690$ & $0.1890$ & $0.5735$\\
after collision & $718$ & $0.1911$ & $0.5692$\\
 \end{tabular}
 \end{center}
 \end{table}

To investigate the details of the structural change, we also
calculate the distribution of $|q_{6m}(i)|^2$ defined by Eq. (B1).
To construct these distributions, we use the particle data 
at the final instant of the initial thermalization by the velocity 
scaling method. Because $q_{lm}$ is equal to $(-1)^{m} q^{*}_{lm}$, 
we only investigate the case of $m \ge 0$.

\begin{figure}[hbtp]
\begin{center}
\includegraphics[width=.4\textwidth]{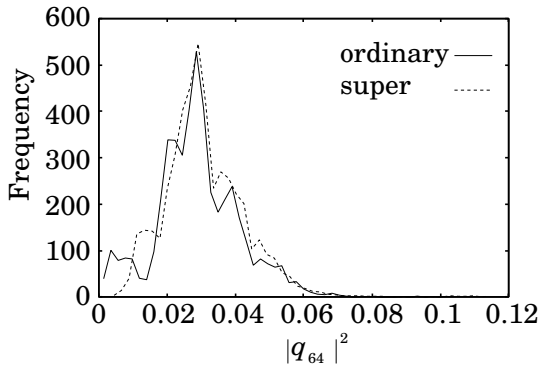}
\end{center}
\caption{
Histogram of $|q_{64}|^{2}$.
}
\label{m-4}
\end{figure}

We present the histogram of $|q_{64}(i)|^{2}$ 
in Fig. \ref{m-4}. The solid curve represents 
the frequency distribution for ordinary rebounds while the broken curve 
is that for super rebounds. 
One can find the slight discrepancy between those two distributions. 
We should mention that the discrepancy is invisible 
for the other $m$, though we do not present the corresponding figures.

\begin{figure}[hbtp]
\begin{center}
\includegraphics[width=.4\textwidth]{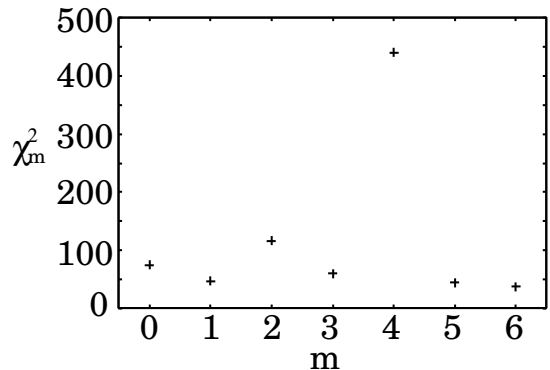}
\end{center}
\caption{
Relationship between $m$ and $\chi_{m}^{2}$ which is defined by Eq. (\ref{chi}).
}
\label{chsq}
\end{figure}

\begin{figure}[hbtp]
\begin{center}
\includegraphics[width=.4\textwidth]{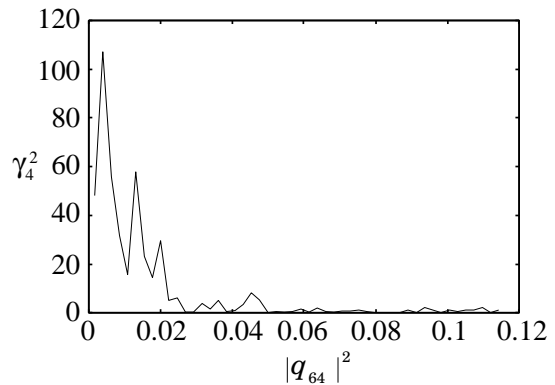}
\end{center}
\caption{
$\gamma_{4}^{2}$ as function of $|q_{64}(i)|^{2}$.
}
\label{rm-4}
\end{figure}

To quantify the discrepancy between those two distributions, 
we introduce $\chi_{m}^{2}$ defined by
\begin{equation}
\chi_{m}^{2} = \sum_{j} \frac{(O_{m}(j)-S_{m}(j))^{2}}{O_{m}(j)+S_{m}(j)} 
\equiv \sum_{j}\gamma_{m}^{2}(j),\label{chi}
\end{equation} 
where $O_{m}(j)$ and $S_{m}(j)$ are the frequencies of $|q_{6m}(i)|^{2}$ 
in the $j$-th bin for ordinary and super rebounds, respectively. 
We calculate $\chi_{m}^{2}$ for each value of $m$ 
to investigate the relationship between $m$ and $\chi_{m}^{2}$. 
Figure \ref{chsq} shows the relationship between $m$ and $\chi_{m}^{2}$,  
which characterizes the difference between super rebounds and ordinary rebounds.  
This figure shows that the discrepancy between the distributions of 
$|q_{64}(i)|^{2}$ 
is remarkably large compared to other values of $m$. 
In addition, we show the relationship between $\gamma_{4}^{2}(j)$ introduced 
in  Eq.(\ref{chi}) and $|q_{64}(i)|^{2}$ in Fig. \ref{rm-4}, 
where a remarkable difference can be found in the range of 
$|q_{64}(i)|^{2} \le 0.02$.


\section{Discussion}

Let us discuss our results. In the previous section, 
we have demonstrated that the structural difference of clusters 
can be found between the super and the ordinary rebounds. 
Here we investigate the potential energy of the characteristic 
local structures included in the clusters 
which induce the super rebounds. 

\begin{figure}[hbtp]
\begin{center}
\includegraphics[width=.2\textwidth]{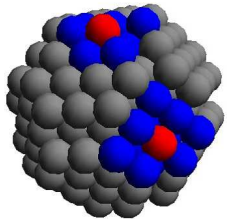}
\end{center}
\caption{
(Color online) Cluster which induces super rebounds. Red (medium grey) particles show the 
``atoms'' with the order 
$1.1643 \times 10^{-2} \le |q_{64}(i)|^{2} \le 1.7425 \times 10^{-2}$ while blue (dark grey) particles 
are neighboring ``atoms''. 
}
\label{nc}
\end{figure}

Focusing on the noticeable peaks in Fig. \ref{rm-4}, 
we select the ``atoms'' with the order in the range, 
$1.1643 \times 10^{-2} \le |q_{64}(i)|^{2} \le 1.7425 \times 10^{-2}$,  
from $20$ clusters which induce the super rebounds. 
The reason why we choose the range of $|q_{64}(i)|^2$ is as follows. 
The characteristic peaks in Fig. \ref{rm-4} are characterized 
by $|q_{64}(i)|^2=(1.4534\pm 0.2891)\times 10^{-2}$, 
so that we pick up ``atoms" in the above range. 
For each of the selected ``atoms'', we define the neighboring ``atoms" 
which are located within the distance of $1.6\sigma$. 
We define a local structure by the collection of the selected atom and 
its neighboring ``atoms''. 
Figure \ref{nc} is a snapshot of a cluster including two local structures, 
each of which is centered by the atom which has the order (red or medium grey). 
This figure shows that the local structures are located on the surface 
of the cluster. Moreover, the local structures seem to be unstable due to 
the less-ordered orientation of ``atoms'', which may lead to the high potential energy.

Next, we calculate the averaged potential energy of those local structures. 
For the purpose, we calculate the potential energy of each atom in the cluster 
with the cutoff $3.5\sigma$ and sum up the potential energy 
over all the ``atoms'' in a local structure to obtain its potential energy. 
Among $20$ clusters which induce super rebounds, 
we find $372$ ``atoms'' to satisfy 
$1.1643 \times 10^{-2} \le |q_{64}|^2 \le 1.7425 \times 10^{-2}$, 
by which we calculate the averaged potential energy of one local structure. 
From our calculation,  the averaged potential energy 
per one local structure becomes $-97.827 \epsilon$. 

 For comparison, we focus on the potential energy of the local structure which 
belongs to the highest peak in Fig. \ref{m-4} 
because we find same number of the local structures 
both in super rebounds and in ordinary rebounds.  
The local structure in this case is defined by  the collection of 
the ``atoms'' with the order around the peak, 
$2.8 \times 10^{-2} \le |q_{64}|^{2} \le 3.0 \times 10^{-2}$,  
and its neighboring ``atoms'' in the clusters which induce 
the super and the ordinary rebounds.  
From the calculation with $482$ local structures in the clusters 
which induces super rebounds, we obtain the averaged potential energy 
per one local structure as $-104.814 \epsilon$. In the case of the 
clusters which induce ordinary rebounds,  the averaged potential energy 
becomes $-104.202\epsilon$ from the calculation of $512$ local structures.  

From these results, the clusters for the super rebound has 
more metastable local structures with high potential energy rather than the clusters 
for the ordinary rebound. Thus, we conjecture that a part of the 
high potential energy is transferred to the macroscopic degrees of freedom 
during the super-rebound collision. 
Moreover, we can also conjecture that the decreases of temperature and entropy after collision are 
connected with the decrease of the potential energy. 
In this paper, we focus on the relationship between the local structure and 
the super rebound. It is also important to understand  the role of macroscopic 
structural change. The macroscopic deformation of nanoclusters 
after the collision will be reported elsewhere.\cite{murakami}




\section{Conclusion}
In this paper, we have performed the molecular dynamics 
simulation of colliding nanoclusters to study 
the origin of super rebounds. 
Through the investigation of the thermodynamic properties, 
we have found that the decreases of the temperature 
and the entropy are observed in one of the binary clusters when the super rebound 
is observed. This may be attributed to the biased distribution 
of the local bond order parameters after the collision which is caused by 
the initial metastable configuration of ``atoms''. 

Through the investigation of the local bond order parameters, 
we have found the discrepancy of the distributions of the order $|q_{64}(i)|^{2}$ 
between the super and the ordinary rebounds. 
The averaged potential energy of the local structure which belongs 
to the characteristic peak in the  $|q_{64}(i)|^{2}$ distribution is larger 
than the typical local structures which are abundant in our cluster.  Thus, 
the energy transfer from the local structure to the macroscopic degree of freedom 
accompanied by the structural change during collision play an important role 
in super rebounds.  Moreover, the change of entropy associated with the change 
in the internal state of order after the collision is the key point 
for understanding the emergence of the super rebounds. 
Further investigation of the relationship between the local structural change 
and the entropy decrease will be one of our future tasks. 

\acknowledgments
We would like to thank N.~V.~Brilliantov, T. Kawasaki, S. Takesue, 
R. Murakami, and K. Saitoh for their valuable comments. Also, HK would like to 
thank Y. Wang and W. Lechner who gave him many advises for calculation of 
order parameters and how to use their calculation codes. 
Parts of numerical computation in this work were carried out 
at Yukawa Institute Computer Facility. 
This work was supported by the Grant-in-Aid for the Global COE Program 
``The Next Generation of Physics, Spun from Universality and Emergence" 
from the Ministry of Education, Culture, Sports, 
Science and Technology (MEXT) of Japan. 

\vspace{3mm}
\appendix
\section{
Calculation of entropy
}\label{appA}
In this appendix, we explain the method to calculate the entropy 
in our system. Let us assume that the collision of C$^{p}$ onto 
C$^{t}$ does the work $\delta W$ and changes the internal energy of 
C$^{t}$. 
Because each cluster is an isolated system, we assume 
that the change in heat $\delta Q$ of C$^{t}$ is determined by 
the first law of thermodynamics,
\begin{equation}
\delta Q = dE + \delta W,  
\end{equation}
where $dE$ is the change in the internal energy.

The internal energy $E$ is calculated as 
\begin{equation}
E = \sum_{i} \frac{{\bf p}_{i}^{2}}{2 m} + 
\frac{1}{2} \sum_{i} \sum_{j \ne i} U(r_{ij}),
\end{equation}
where ${\bf p}_{i}$, $m$ are the relative momentum of an atom 
in C$^{t}$ against the center of mass and the mass of an atom, 
respectively. $U(r_{ij})$ is the potential energy between the 
``atoms'' $i$ and $j$ in C$^{t}$. 

\begin{figure}[hbtp]
\begin{center}
\includegraphics[width=.3\textwidth]{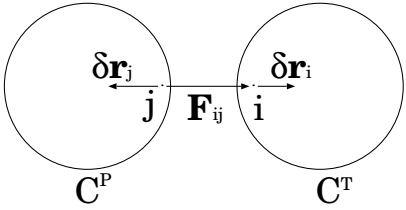}
\end{center}
\caption{
Interaction between an atom in C$^{t}$ and an atom in C$^{p}$.
}
\label{app_fig1}
\end{figure}

The amount of work $\delta W(t)$ acting on C$^{t}$ is defined as follows. 
Let us define the work done by the atom $j$ of C$^{p}$ 
to the atom $i$ of C$^{t}$ (Fig. \ref{app_fig1}). 
Assuming that the displacements of the ``atoms'' are caused by the 
interactive force ${\bf F}_{ij}$, the amount of work 
$\delta W_{ij}(t)$ done by the atom $j$ to the atom $i$ 
during an interval $dt$ may be written as 
\begin{equation}
\delta W_{ij}(t) = {\bf F}_{ij} \cdot \left[
{\bf r}_{ij}(t+dt) -  {\bf r}_{ij}(t) 
\right],
\end{equation}
where ${\bf r}_{ij}(t) = {\bf r}_{i}(t) - {\bf r}_{j}(t)$. 
Thus, the work $\delta W(t)$ done by C$^{p}$ to C$^{t}$ 
may be expressed as 
\begin{equation}\label{work}
\delta W(t) = \frac{1}{2} \sum_{i} \sum_{j} \delta W_{ij}(t).
\end{equation}

Using these quantities, 
we define the entropy difference $\Delta S(t) = S(t) - S(0)$ as
\begin{equation}\label{entr}
\Delta S(t) = \int_{0}^{t} \frac{\delta Q(t^{'})}{T(t^{'})}.
\end{equation}
Here we use the kinetic temperature Eq.(\ref{temp}) as $T$. 
To calculate $\Delta S(t)$ by Eq.(\ref{entr}), 
we first smooth the temperature and the heat as
\begin{eqnarray}
\tilde{T}(t) &=& \int_{t-\Delta t}^{t+\Delta t} T(t^{'}) dt^{'}\\
\delta \tilde{Q}(t) &=& \int_{t-\Delta t}^{t+\Delta t} \delta Q(t^{'}) dt^{'}\\
\end{eqnarray}
with $\Delta t= 0.565 \sigma/\sqrt{\epsilon/m}$ 
because those data change rapidly against time. 
By using those smoothed parameters, the entropy change of 
C$^{t}$ may be rewritten as 
\begin{eqnarray}\label{entropy}
\Delta{S}(t) = \int_{0}^{t} \frac{\delta \tilde{Q}(t^{'})}{\tilde{T}(t^{'})},
\end{eqnarray}
where we adopt the simple trapezoidal rule for the evaluation of the integral.

\section{Calculation of order parameter}\label{appB}
In this appendix, we introduce Steinhardt's order parameter 
to characterize the structure of our model
\cite{steinhardt,lechner,wolde}. 
Let us assume that an atom $i$ in a crystalline structure 
is surrounded by $N_{b}(i)$ ``atoms'' within the cutoff length $r_{c}$. 
That is, the distance $r_{ij}$ between the $i$-th atom and 
one of the neighboring ``atoms'' $j$ is less than the cutoff 
length $r_{c}$. In our analysis, we adopt $r_{c}=1.6 \sigma$. 

First, for each atom, we calculate the average of spherical 
harmonics which depends on $r_{ij}$ as follows:
\begin{equation}
q_{lm}(i) = \frac{1}{N_{b}(i)} 
\sum_{j=1}^{N_{b}(i)} Y_{lm}(r_{ij}).
\end{equation}
Next, we introduce $q_{l}(i)$ according to the following definition:
\begin{equation}
q_{l}(i) = \sqrt{\frac{4\pi}{2l+1} \sum_{m=-l}^{l} |q_{lm}(i)|^{2}}.
\end{equation}
Instead of $q_{l}(i)$, we use the time-averaged value defined by 
\begin{eqnarray}\label{eqB3}
Q_{l}(i) = \frac{1}{\tau_{\alpha}} \int_{t_{0}}^{t_{0}+\tau_{\alpha}} q_{l}(i) dt
\label{Qli}
\end{eqnarray}
with the time interval $\tau_{\alpha}$.

Each crystalline structure, 
such as a body-centered cubic (bcc) and a face-centered cubic (fcc), 
has a characteristic distributions of those parameters. 
For example, the distributions of $q_{4}$ and $q_{6}$ in a fcc structure shows 
two peaks while the distributions in other structure has a single peak 
in a Lennard-Jones system.\cite{lechner} 
The mean values of those distributions 
are summarized in Table \ref{tabl3}.\cite{wolde}


 \begin{table}
 \caption{\label{tabl3}
Bond orientational order parameters for fcc, bcc, and hcp crystals\cite{wolde}.
}
 \begin{center}
 \begin{tabular}{lcc}
 & $q_{4}$ & $q_{6}$\\
\hline
fcc & $0.191$ &   $0.575$\\
bcc & $0.036$ &   $0.511$\\
hcp & $0.097$ &   $0.485$
 \end{tabular}
 \end{center}
 \end{table}


\end{document}